\documentclass[twocolumn,secnumarabic,amssymb, nobibnotes, aps, prd]{revtex4-2}

\usepackage{graphicx, bm}
\begin{document}

\title{Frame Superposition Cluster: The method to derive the transition matrices in high accuracy}

\author{Hikaru Wakaura}%
\email[Quantscape: ]{
hikaruwakaura@gmail.com}
\affiliation{QuantScape Inc., 4-11-18, Manshon-Shimizudai, Meguro, Tokyo, 153-0064, Japan}

\author{Takao Tomono}
\email[Toppan Inc.: ]{takao.tomono@ieee.org}
\affiliation{ Digital Innovation Div. Toppan Inc. 1-5-1, Taito, Taito, Tokyo, 110-8560, Japan}

\date{July 2021}%

\begin{abstract} 
Variational Quantum Eigensolver (VQE) method is the way to derive the wave functions from their eigenenergies using quantum computers. But, the methods to utilize the superposition states between them haven't been developed yet. The method to calculate the off-diagonal element of observables between two states derived by VQE method requires large numbers of gate operations for Noisy-Intermediate-Scale-Quantum (NISQ) devices. We propose the novel method to derive the superposition state between the states derived by VQE method. We call it Frame Superposition Cluster (FSC) method. Using the method, we confirmed that dipole transition moment could be calculated with the highest accuracy compared to other methods. 
\end{abstract}

\maketitle

\section{Introduction}\label{1}
Since introduced by Aspuru-Guzik et al. in 2011 \cite{doi:10.1146/annurev-physchem-032210-103512}, the Variational Quantum Eigensolver (VQE) has been studied and improved by various groups. Cloud quantum computing with 5 qubits by IBM initiated the movement to apply the calculation of VQE method on various molecules \cite{PhysRevX.8.011021, 2017Natur.549..242K} and quantum systems, which eventually boost its further development. Today, various types of VQE methods are brought out to the world. For example, Subspace-Search VQE (SSVQE) \cite{PhysRevResearch.1.033062} can calculate the multiple energy levels coincide and Multiscale-Contracted VQE (MCVQE) \cite{2019PhRvL.122w0401P} calculates the ground and single electron excited states by diagonalizing Configuration Interaction State (CIS) Hamiltonian. Adaptive VQE \cite{2019NatCo..10.3007G} and Deep-VQE \cite{2020arXiv200710917F} are also proposed. The essential procedure of quantum circuit learning has taken advantage of VQE \cite{PhysRevA.98.032309}.
In parallel to this movement, quantum hardware has been improved concerning both the number of qubits and quantum volume. The fidelity of qubits skyrocketed last year. Both Honeywell \cite{2020PhRvR...2a3317B} and Ion-Q \cite{IonQ2020} updated the record of quantum volume twice and the record of this is four million achieved by the quantum computer of Ion-Q. Recently, the Institute of Science in China has achieved quantum supremacy by photonic quantum computer \cite{Zhong1460}. 

However, the methods of using derived states have not been developed except SSVQE method. The method to calculate the transition matrices has been established in 2018 by QunaSys\cite{2020arXiv200211724I} even it is low accuracy and hard to use in NISQ devices. Moreover, there are no practical methods to make superposition states between each state. We propose the new method to make a given superposition state between states derived by VQE method using the unitary coupled cluster (UCC) method. We call it Frame Superposition Clusters (FSCs). This cluster transit the given state to other states by acting on given times. We revealed that it can calculate the dipole moments with higher accuracy. 

The following organizations of this paper are as follows. Chapter \ref{2} is described the detail of our method and FSCs. Chapter \ref{3} indicates the result of our work calculation on dipole moments of hydrogen and helium hydride molecules using FSCs. Chapter \ref{4} is the conclusion of our works.

\section{Method}\label{2}

In this section, we describe how to make the FSCs. FSCs transit the quantum states by VQE method to other states. At first, we describe the method to make the states by VQE method. The state of VQE method is derived by optimizing,

\begin{eqnarray}
E_j(\bm{\theta^j})&=&\langle \Phi_{ini} \mid U^\dagger(\bm{\theta^j})HU(\bm{\theta^j})\mid \Phi_{ini} \rangle \\ \nonumber
&=&\langle \Phi_j \mid H \mid \Phi_j \rangle \label{E}
\end{eqnarray}.

Then, $E_j$ is the trial eigenenergy of $j^{th}$ state and $U$ is the operators to make the given superposition state, respectively. $U$ includes the translated Unitary Coupled Cluster(UCC) to make the superposition state and the propagator of the operators of hamiltonian as Pauli words\cite{doi:10.1021/acs.jctc.8b00450}. 

There are many types of cluster to prepare the quantum states. For example, Deep Multiscale Entanglement Renormalization Amsatz (DMERA)\cite{2017arXiv171107500K}, Trotterized version of Adiabatic State Preparation (TASP)\cite{PhysRevA.92.042303}, Qubit Cluster Coupling (QCC)\cite{2018arXiv180903827R} and some various low depth ansatz\cite{2017Natur.549..242K}.
We apply Unitary Coupled Cluster of Single and Double (UCCSD) on the condition of single and double excitation terms\cite{2018PhRvA..98b2322B} because it has the highest accuracy even time for calculation is not short. The depth of UCCSD and hamiltonian are both 2.
To make the excited state, Variational Quantum Deflation (VQD)\cite{2018arXiv180508138H} term and constraint terms\cite{doi:10.1021/acs.jctc.8b00943} are also minimized. All states are derived from ground state to highest excited states one after another. Off diagonal elements of observable between two trial states can be calculated. It can be  expressed as,

\begin{eqnarray} \nonumber
\langle \Phi_j \mid H \mid \Phi_k \rangle &=& a + ib \\ \nonumber
\langle + \mid H \mid + \rangle - \frac{1}{2}(\langle \Phi_j \mid H \mid \Phi_j \rangle + \langle \Phi_k \mid H \mid \Phi_k \rangle)&=&a \\ \label{od}
\langle y+ \mid H \mid y+ \rangle - \frac{1}{2}(\langle \Phi_j \mid H \mid \Phi_j \rangle + \langle \Phi_k \mid H \mid \Phi_k \rangle)&=&b \\ \nonumber
\end{eqnarray}.

Then, $\mid + \rangle$ and $\mid y+ \rangle$ are equally superposition state of $j^{th}$ and $k^{th}$ state and equally superposition state with the state phase $\pi/2$, respectively. They must be calculated by some unique technics.
FSCs are the cluster operators that transit the given states to other states by acting n times. For example, they are expressed as,

\begin{equation}
U_{j\rightarrow k}^n\mid \Phi_j \rangle = \mid \Phi_k \rangle
\end{equation}. 

They are UCCSD same as that of VQE. They are made by variational method same as VQE. $\mid+\rangle$ and $\mid y+\rangle$ are made by same process by following equation,

\begin{equation}
\mid + \rangle = U_{j\rightarrow k}^{n/2}\mid \Phi_j \rangle,~\mid y+ \rangle=U_{+\rightarrow~-}^{n/2}\mid + \rangle
\end{equation}.

By minimizing a and b in eq. \label{od}, we optimize the FSCs and their variables to calculate the off-diagonal elements. Evaluation function is as follows,

\begin{equation}
F=a(H)+b(H)+H_{const}
\end{equation}.

Then, $H_{const}$ is the constrain term for FSCs expressed as $\mid \langle \Phi_k \mid H \mid \Phi_k \rangle - \langle  \Phi_j\mid U_{j\rightarrow k}^{n\dagger}H U_{j\rightarrow k}^{n}\mid \Phi_j \rangle \mid$. We optimized the each state by VQE using Broyeden-Flecher-Goldferb-Shanno(BFGS) method for optimizer and optimized the FSCs using Sequential Least Squares Programing(SLSQP). The depth of $U_{j\rightarrow k}^{n/2}$ and $U_{+\rightarrow -}^{n/2}$ are both 4. All datas are calculated by blueqat SDK and openfermion\cite{2017arXiv171007629M}.

\section{result}\label{3}
In this section, we present our result of the calculation on dipole moment between each state of hydrogen and helium hydride molecules in case diatomic bond length is 0.7($\AA$) calculated by FSCs. We also compare our result to that of the method of calculation on VQE and SSVQE. $n=2$ for all cases. As for hydrogen molecules, FSC can derive dipole moments with high accuracy. As shown in Table. \ref{Hvqedp}, and \ref{Hssvqedp}, derived dipole moments by VQE and SSVQE are not negligible. However, derived dipole moments by VQE and FSCs are all negligible as shown in Table. \ref{Hfscdp}. Dipole moments of hydrogen molecules are zero apparently from their shape. Low accuracy of each energy level is supposed to affect to low accuracy of dipole moments. As shown in Table. \ref{Hvqedp} and \ref{Hene}, dipole moment between singlet and doubly excited state is larger than others and the common difference between calculated and exact energy(log error) of the doubly excited state is above -2. However, shown in Table. \ref{Hene}, FSCs can derive transition matrices regardless of the accuracy of each state. Table. \ref{Hdiff} is the table of energy differences between transmitted initial state to destination state by FSCs and destination states. They are all nearly zero between each state. Hence, the FSCs transits each state correctly.
The result of the calculation on helium hydride molecules is also high accuracy in the case calculated by FSCs. As shown in Table. \ref{HeHvqedp} and \ref{HeHssvqedp}, dipole moments calculated by VQE and SSVQE are not negligible. The result by VQE is greater than that of SSVQE because the first ground state deviates from its exact value as shown in Table. \ref{HeHvqedp} and \ref{HeHssvqedp}. The result by FSCs is negligible same as exact values between two excited state -1.400$\times$10$^{-16}$ (Debye) regardless of low accuracy of energy levels as shown in Table. \ref{HeHfscdp} and \ref{HeHene}. The energy differences between initial states changed to destination state and destination states are negligible as shown in Table. \ref{HeHdiff} same as hydrogen molecules.

\begin{table}
\caption{
The transition dipole moment of ground, triplet, singlet and doubly excited states on hydrogen molecules calculated by VQE with VQD. The unit of dipole moment is Debye.
}\label{Hvqedp}

\begin{tabular}{c|cccc} \hline \hline
&Ground&Triplet&Singlet&Doubly \\ \hline
Ground&0&0.074&0.0285&0.0191 \\
Triplet&---&0&0.0773&0.0302 \\
Singlet&---&---&0&47.3433 \\
Doubly&---&---&---&0 \\ \hline
\end{tabular}
\end{table}

\begin{table}
\caption{
The transition dipole moment of ground, triplet, singlet and doubly excited states on hydrogen molecules calculated by SSVQE. The unit of dipole moment is Debye.
}\label{Hssvqedp}

\begin{tabular}{c|cccc} \hline \hline
&Ground&Triplet&Singlet&Doubly \\ \hline
Ground&0&2.6741&1.7808&0.9556 \\
Triplet&---&0&1.1598&0.2976 \\
Singlet&---&---&0&0.8157 \\
Doubly&---&---&---&0 \\ \hline

\end{tabular}
\end{table}

\begin{table}
\caption{
The transition dipole moment of ground, triplet, singlet and doubly excited states on hydrogen molecules calculated by VQE with VQD used for FSCs. The unit of dipole moment is Debye.
}\label{Hfscdp}

\begin{tabular}{c|cccc} \hline \hline
&Ground&Triplet&Singlet&Doubly \\ \hline
Ground&0&3.394$\times$10$^{-11}$&2.930$\times$10$^{-11}$&2.819$\times$10$^{-11}$ \\
Triplet&---&0&3.599$\times$10$^{-11}$&3.492$\times$10$^{-11}$ \\
Singlet&---&---&0&3.142$\times$10$^{-11}$ \\
Doubly&---&---&---&0 \\ \hline
\end{tabular}
\end{table}

\begin{table}
\caption{
The energy levels and their log errors of ground, triplet, singlet and doubly excited states on hydrogen molecules calculated by VQE with VQD, SSVQE and VQE with VQD used for FSCs. The unit of energy is Hartree.
}\label{Hene}
\begin{tabular}{c|cc|cc|cc} \hline \hline
&VQD& &SSVQE& &FSC& \\
&Energy&Log error&Energy&Log error&Energy&Log error \\ \hline
Ground&-1.1362&-10.8639&-1.1229&-1.8781&-1.136&-10.482 \\
Triplet&-0.4785&-9.5757&-0.4736&-2.311&-0.479&-9.485 \\
Singlet&-0.1285&-2.0964&-0.1228&-2.6359&-0.122&-2.795 \\
Doubly&0.5107&-1.1388&0.5448&-1.4141&0.569&-1.848 \\ \hline
\end{tabular}
\end{table}

\begin{table*}
\caption{
The $H_{const}$ between ground, triplet, singlet and doubly excited states on hydrogen molecules calculated by VQE with VQD used for FSCs. 1, 2, 3 and 4 indicates ground, triplet, singlet and doubly excited states, respectively.
}\label{Hdiff}
\begin{tabular}{c|cccccc} \hline \hline
(i, j)&(1,2)&(1,3)&(1,4)&(2,3)&(2,4)&(3,4) \\ \hline
$H_{const}$&2.18 $\times$10$^{-13}$&3.74$\times$10$^{-2}$&3.85$\times$10$^{-5}$&3.85$\times$10$^{-5}$&5.13$\times$10$^{-13}$&1.77$\times$10$^{-3}$ \\ \hline

\end{tabular}
\end{table*}

\begin{table}
\caption{
The transition dipole moment of doubly ground and excited states on helium hydride molecules calculated by VQE with VQD. The unit of dipole moment is Debye.
}\label{HeHvqedp}

\begin{tabular}{c|cccc} \hline \hline
&Ground1&Ground2&Excited1&Excited2\\ \hline
Ground1&0&16.7101&15.1649&16.6579 \\
Ground2&---&0&4.0421&11.3985 \\
Excited1&---&---&0&16.2443 \\
Excited2&---&---&---&0\\ \hline

\end{tabular}
\end{table}

\begin{table}
\caption{
The transition dipole moment of doubly ground and excited states on helium hydride molecules calculated by SSVQE. The unit of dipole moment is Debye.
}\label{HeHssvqedp}

\begin{tabular}{c|cccc} \hline \hline
&Ground1&Ground2&Excited1&Excited2 \\ \hline
Ground1&0&2.6741&1.7808&0.9556 \\
Ground2&---&0&1.1598&0.2976 \\
Excited1&---&---&0&0.8157 \\
Excited2&---&---&---&0 \\ \hline

\end{tabular}
\end{table}

\begin{table}
\caption{
The transition dipole moment of doubly ground and excited states on helium hydraide molecules calculated by VQE with VQD used for FSCs. The unit of dipole moment is Debye.
}\label{HeHfscdp}

\begin{tabular}{c|cccc} \hline \hline
&Ground1&Ground2&Excited1&Excited2 \\ \hline
Ground1&0&1.433&1.433&1.433 \\
Ground2&---&0&1.133$\times$10$^{-5}$&0.852$\times$10$^{-3}$ \\
Excited1&---&---&0&0.841$\times$10$^{-3}$ \\
Excited2&---&---&---&0 \\ \hline

\end{tabular}
\end{table}

\begin{table}
\caption{
The energy levels and their log errors of doubly ground and excited states on helium hydride molecules calculated by VQE with VQD, SSVQE and VQE with VQD used for FSCs.
}\label{HeHene}
\begin{tabular}{c|cc|cc|cc} \hline \hline
&VQD&&SSVQE&&FSC& \\
&Energy&Log error&Energy&Log error&Energy&Log error \\ \hline
Ground1&-2.396&-0.2695&-2.9335&-3.9174&-2.396&-0.2695 \\
Ground2&-2.9337&-10.1565&-2.9333&-3.4203&-2.9337&-10.247 \\
Excited1&-1.8201&-4.9324&-1.8202&-3.8121&-1.8201&-5.9984 \\
Excited2&-1.8201&-8.1579&-1.8195&-3.2007&-1.8201&-4.4367 \\ \hline

\end{tabular}
\end{table}

\begin{table*}
\caption{
The $H_{const}$ between ground, triplet, singlet and doubly excited states on hydrogen molecules calculated by VQE with VQD used for FSCs. 1, 2, 3 and 4 indicates ground1, ground2 excited1 and excited2 states, respectively.
}\label{HeHdiff}
\begin{tabular}{c|cccccc} \hline \hline\ 
(i, j)&(1,2)&(1,3)&(1,4)&(2,3)&(2,4)&(3,4) \\ \hline
$H_{const}$&5.37&3.92$\times$10$^{-10}$&9.38$\times$10$^{-11}$&5.91$\times$10$^{-4}$&3.80$\times$10$^{-4}$&3.63$\times$10$^{-4}$ \\ \hline

\end{tabular}
\end{table*}

\section{Conclusion}\label{4}

From these results, it is revealed that FSCs can derive the equally superposition state and equally superposition state with the phase of given two states in high accuracy regardless of the accuracy of the two states. This result is don't depend on the kind of molecules in the diatomic molecules. However, FSCs take more time than the method in the paper. Improving the time for optimization is the next problem. To derive the FSCs and calculate the transition matrices on complex molecules such as diazine is also a problem. We demonstrated that FSCs could calculate the dipole moment by the far less depth of the circuit.

\bibliographystyle{apsrev4-2}
\bibliography{temp6}

\end{document}